\documentclass[%
 prl, twocolumn,
superscriptaddress,
 amsmath,amssymb,
 aps,
]{revtex4-1}

\usepackage{graphicx}
\usepackage{dcolumn}
\usepackage{bm}
\usepackage{xcolor}
\usepackage{xr}
\usepackage{bm}
\usepackage{hyperref}
\usepackage{enumitem}
\usepackage{accents}
\usepackage{comment}
\hypersetup{colorlinks,linkcolor={blue!90!black},citecolor={blue!90!black},urlcolor={blue!90!black}}

\DeclareMathAlphabet\mathbfcal{OMS}{cmsy}{b}{n}
\DeclareMathOperator\erfc{erfc}

\usepackage{amsmath}
\usepackage{mathtools}
\usepackage{euscript}

\begin{document}

\title{How do Cicadas Emerge Together?\\
Thermophysical Aspects of Their Collective Decision-Making}

\author{Raymond E. Goldstein}
  \email{R.E.Goldstein@damtp.cam.ac.uk}
  \affiliation{Department of Applied Mathematics and Theoretical 
Physics, Centre for Mathematical Sciences,\\ University of Cambridge, Wilberforce Road, Cambridge CB3 0WA, 
United Kingdom}
\author{Robert L. Jack}
  \email{rlj22@cam.ac.uk}
\affiliation{Department of Applied Mathematics and Theoretical 
Physics, Centre for Mathematical Sciences,\\ University of Cambridge, Wilberforce Road, Cambridge CB3 0WA, 
United Kingdom}
\affiliation{Yusuf Hamied Department of Chemistry, University of Cambridge, Lensfield Road, Cambridge CB2 1EW, United Kingdom}
  \author{Adriana I. Pesci}
\email{A.I.Pesci@damtp.cam.ac.uk}
\affiliation{Department of Applied Mathematics and Theoretical 
Physics, Centre for Mathematical Sciences,\\ University of Cambridge, Wilberforce Road, Cambridge CB3 0WA, 
United Kingdom}
\date{\today}

\begin{abstract}
Certain periodical cicadas exhibit life cycles with durations of $13$ or $17$ years, and it is now generally
accepted that such large prime numbers arise evolutionarily to avoid synchrony with predators.  Less well explored
is the question of {\it how}, in the face of intrinsic biological and environmental noise, insects within a brood emerge together
in large successive swarms from underground during springtime warming.  Here we consider the decision-making process of underground cicada nymphs experiencing random but spatially-correlated thermal microclimates like those in nature.  Introducing short-range communication between insects leads to a model 
of consensus building that 
maps on to the statistical physics of 
an Ising model with a quenched, spatially correlated random magnetic field and annealed site dilution, which displays the kinds of
collective swarms seen in nature.
\end{abstract}

\maketitle

The synchronized spring-time emergence from underground of cicadas of the genus {\it Magicicada}
has been the subject of detailed entomological field studies for over a century \cite{Simon2022}.
From work documenting the geographic 
distribution of emergences of $13-$ or $17-$year species \cite{Marlatt1907}, 
to studies of their underground developmental 
stages \cite{Lloyd1966,Heath1968,Maier1996}, it is understood
that any given brood (group emerging in a particular year) exhibits two types of
synchrony; (i) essentially all members emerge precisely in year $13$ or $17$, and 
(ii) they do so when the local soil temperature crosses a threshold in that particular year \cite{Heath1968}.  

These observations motivated numerous studies in theoretical population biology 
to understand the reasons {\it why} large prime number periods have been selected by evolution, but
far fewer studies explaining {\it how} the two levels of synchrony are achieved.  
For prime number selection, the hypothesis \cite{AlexanderMoore1962,WhiteLloyd1975} that 
limited environmental carrying
capacity and predation pressure are responsible was first captured in a mathematical model by
Hoppensteadt and Keller \cite{HoppensteadtKeller}.  Later models elucidated mechanisms by which single broods occupy disjoint 
areas \cite{Blackwood2018,Machta2019,Diekmann2020}.

\begin{figure}[b]
\includegraphics[width=0.98\columnwidth]{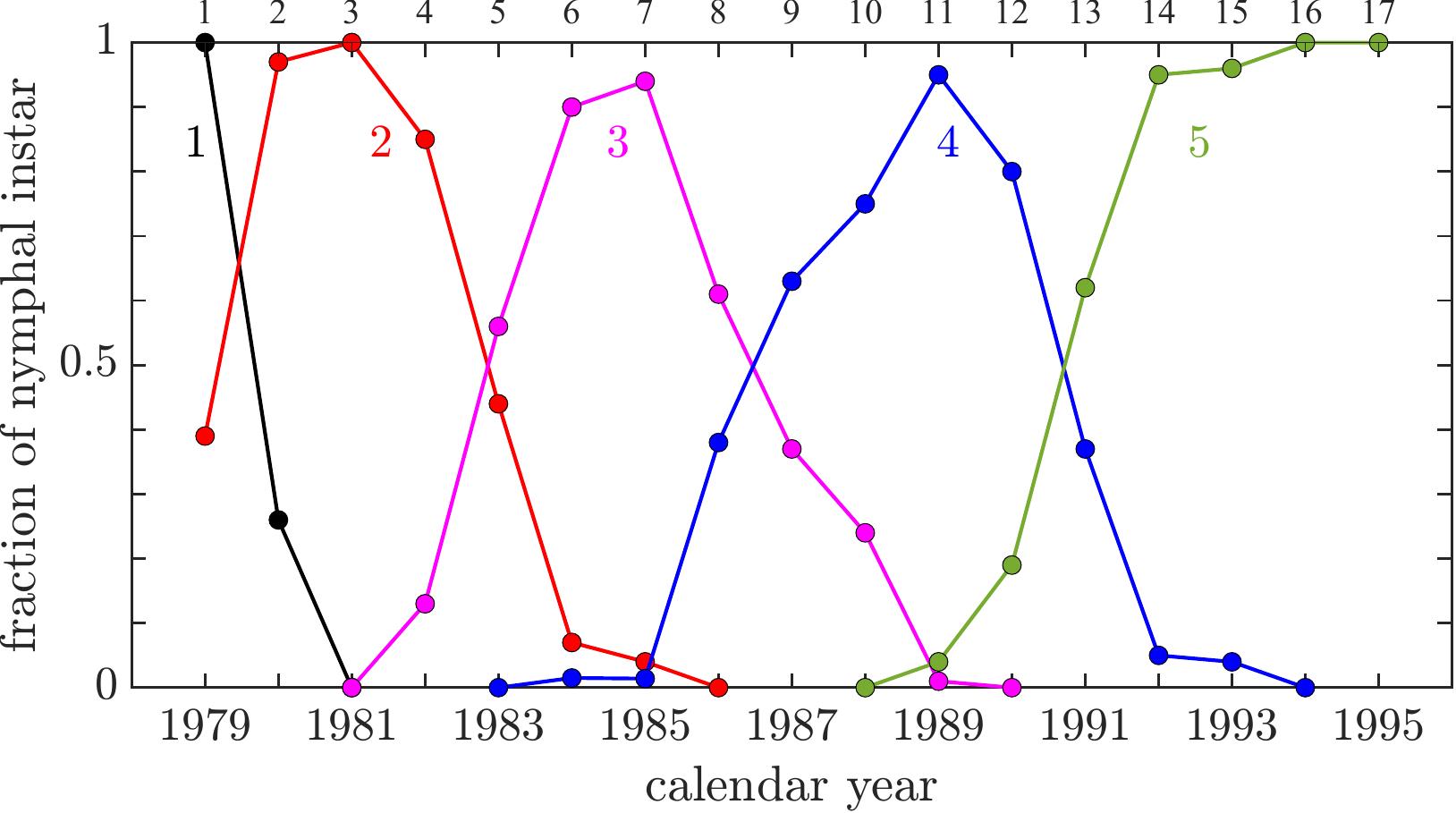}
\caption{Proportion of cicadas in the $5$ instars as a function of time, for one brood.  Adapted from
Ref. \cite{Maier1996}.}
\label{fig1}
\end{figure}

These studies do not address how
a brood recognizes that it is year $17$ (and not, say, $16$) and
then emerges in a sequence of vast swarms throughout several weeks.  The $17$ years spent
underground by nymphas are divided into $5$ {\it instars}, the duration of which exhibits considerable dispersion 
(Fig. \ref{fig1}).   
Despite this
spread, cicadas accurately keep track
of the passage of years while underground.  It is known that after hatching 
the nymphs burrow below ground and obtain nutrients from the xylem in tree roots \cite{Karban1982}.  
They therefore experience the annual seasonal cycles of the trees, as shown
by Karban, et al. \cite{Karban2000}, who artificially altered the cycles in year $15$
to provoke an early emergence, proving that cicadas count cycles and not the passage of time itself.
It is unclear how such accurate counting occurs, but it has been
suggested \cite{Simon2022} that it could involve epigenetic modifications of the
kind observed in long-lived plants like bamboo \cite{bamboo}.
Similar issues arise in flowin order to flower \cite{polycomb}.

\begin{figure*}[t]
\includegraphics[width=1.98\columnwidth]{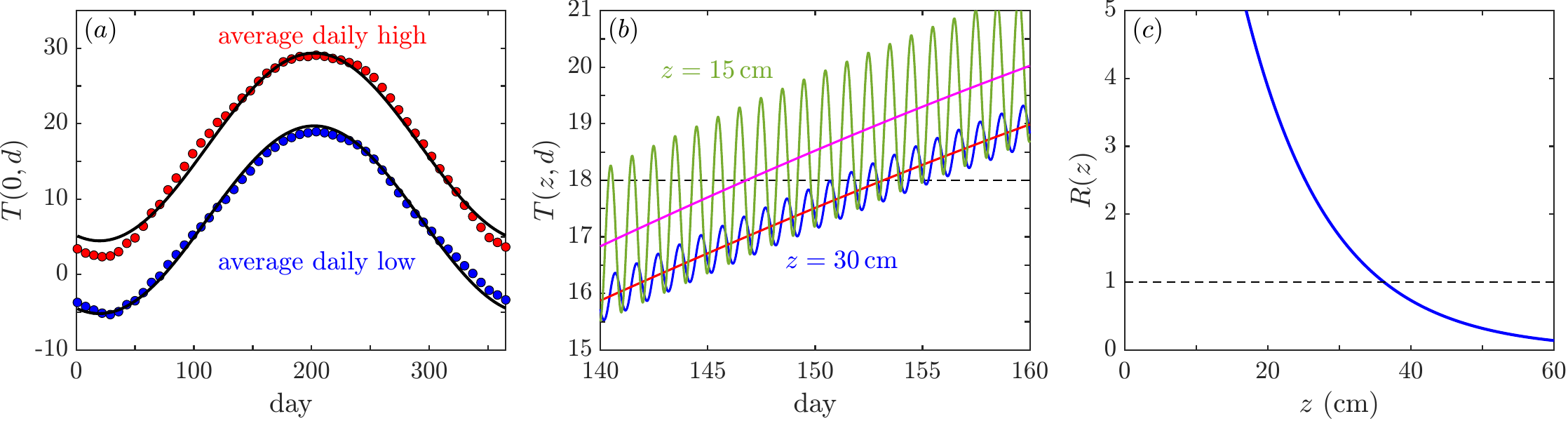}
\caption{Temperature variations. (a) Daily average low and high surface temperatures in Columbus, Ohio, subsampled weekly,
and daily extrema of two-mode approximation \eqref{surface_temp} (black). 
(b) Theoretical average subsurface temperature at two depths near $T_c=18^\circ$C. (c) Noise ratio $R$ in \eqref{noiseratio} versus depth near the crossing day.}
\label{fig2}
\end{figure*}

The issue of swarm emergence in a given year was studied by Heath \cite{Heath1968},
who found that the day $d_c$ of emergence of 17-year cicadas in any given location is strongly
correlated with the local soil temperature reaching the threshold $T_c\simeq 18^\circ$C. 
This conclusion raises the question of how cicadas can emerge
in great swarms in spite of spatially-varying microclimates, their own distribution of body
temperatures on emergence \cite{Heath1968}, and the inherent imprecision of temperature sensing by
the cicadas themselves. 
Here we develop the hypothesis that the thermally-triggered 
synchronized emergence of cicadas arises in part from 
short-ranged communication between nearby underground nymphs that allows for 
collective decision-making.
That cicadas are capable of collective behavior by means of communication is evidenced by their acoustically synchronized above-ground
choruses \cite{SiYuan2009,Sheppard2020}.  
While choruses occur soon after emergence, and it is plausible that the ability to hear underground noise \cite{Maeder2022}
is present earlier, acoustical coupling is but one of several communication mechanisms that may
be operating, and our analysis does not depend on the specific means.  Collective behavior via 
communication \cite{Couzin} is found in many contexts:  bacterial quorum sensing \cite{quorum}, ant foraging  
\cite{Beekman2001,Sasaki2013,Feinerman2018}, and bird flocking \cite{Cavagna2010}.

As in other studies of collective behavior \cite{Michard2005,Bouchaud2013}, our 
model of decision-making  
is a random-field Ising model (RFIM) \cite{Sethna1993,Sethna2005}, in which 
quenched randomness arises from microclimates and spins represent the decision.  We introduce additional site occupancy variables in order to
interpret the simultaneous flipping of many spins (``avalanches") as swarms.  Numerical studies 
of this model
produce swarms like those found in nature.

{\it Thermobiology of burrowing nymphs.}  Newly hatched nymphs burrow to a depth
$z_b\sim\! 30\,$cm that is thought from observations \cite{Heath1968} to 
isolate them from strong diurnal temperature fluctuations.
To put this on a quantitative basis, we consider 
the temperature variations in Ohio, where there is a wealth of data on cicada 
emergence \cite{Heath1968}.  Figure \ref{fig2}(a) shows the average daily low and high temperatures at $2\,$m above ground in Columbus, Ohio \cite{temp_data}.  We take these
to define a suitable average boundary condition $T(0,d)$ for the subsurface 
temperature field $T(z,d)$ with $z$ increasing downward
and $d$ is time measured in days. 
These data can be represented
by a two-term Fourier series corresponding to a superposition of
annual ($a$) and daily ($d$) cycles, 
\begin{equation}
    T(0,d)=\bar{T}-\Delta_a\cos\left(2\pi\nu_a d'\right)
    -\Delta_d\cos\left(2\pi\nu_d d'\right),
    \label{surface_temp}
\end{equation}
where $d'=d-d_0$, with $d_0\simeq 20$ (January 20th) being the day of lowest temperatures, with annual frequency $\nu_a=(1/365)\,$day$^{-1}$, daily frequency
$\nu_d=1\,$day$^{-1}$,  
$\bar{T}=12.1^\circ$C, $\Delta_a=12.4^\circ$C and 
$\Delta_d=4.8^\circ$C.
We assume the underground temperature $T(z,d)$ obeys the diffusion equation
$\partial_d T=D\partial_{zz}T$, for which typical values of the thermal
diffusion constant $D$ are in the range
$(0.8-10)\times 10^{-7}\,$m$^2$/s \cite{soilD}. We adopt the middle of this
range $D\sim 5\times 10^{-7}\,$m$^2$/s$= 432\,$cm$^2$/day.

Introducing the scaled time $t=\nu_d d'$ and
$\epsilon=\nu_a/\nu_d$, Eq. \eqref{surface_temp} implies the subsurface temperature field
\begin{align}
    T(z,t)=\bar{T}&-\Delta_a e^{-z/\ell_a}\cos\left(z/\ell_a-2\pi\epsilon t\right) \nonumber \\
    &-\Delta_d e^{-z/\ell_d}\cos\left(z/\ell_d-2\pi t\right),
    \label{subsurfacetemp}
\end{align}
with penetration lengths $\ell_i=\sqrt{D/\pi\nu_i}$ for $i=a,d$, with values $\ell_a\sim 224\,$cm and $\ell_d\sim 12\,$cm, respectively.
Examining the subsurface temperature field at different depths, as in Fig. 
\ref{fig2}(b), we see that at $z=15\,$cm the within-day oscillations are very large
compared to the change in the mean between successive days, whereas at
$z=30\,$cm the two are comparable. To quantify 
the relative size of these two contributions we define $R(z)$ as
the ratio between the root-mean-square daily fluctuations in temperature and the 
change in the annual trend over
one day.  Since $\epsilon\ll 1$, we approximate $R(z,t)$ as
\begin{equation}
    R(z,t)=-\frac{\Delta_d\, e^{-z\left(\frac{1}{\ell_d}-\frac{1}{\ell_a}\right)}}{2^{3/2}\pi\epsilon \Delta_a\sin\left[z/\ell_a-2\pi\epsilon t\right]}.
    \label{noiseratio}
\end{equation}
Shown in Fig. \ref{fig2}(c), this ratio decreases with depth, crossing below unity 
at the burrowing depth $z_b\sim 30\,$cm. While
fluctuations are attenuated relative to the surface, the thermal noise there 
is comparable to the signal, and thus crossing of the temperature threshold can not be synchronously 
determined by a population of nymphs, buried at a distribution of depths, acting independently.

\begin{figure}[t]
\includegraphics[width=0.98\columnwidth]{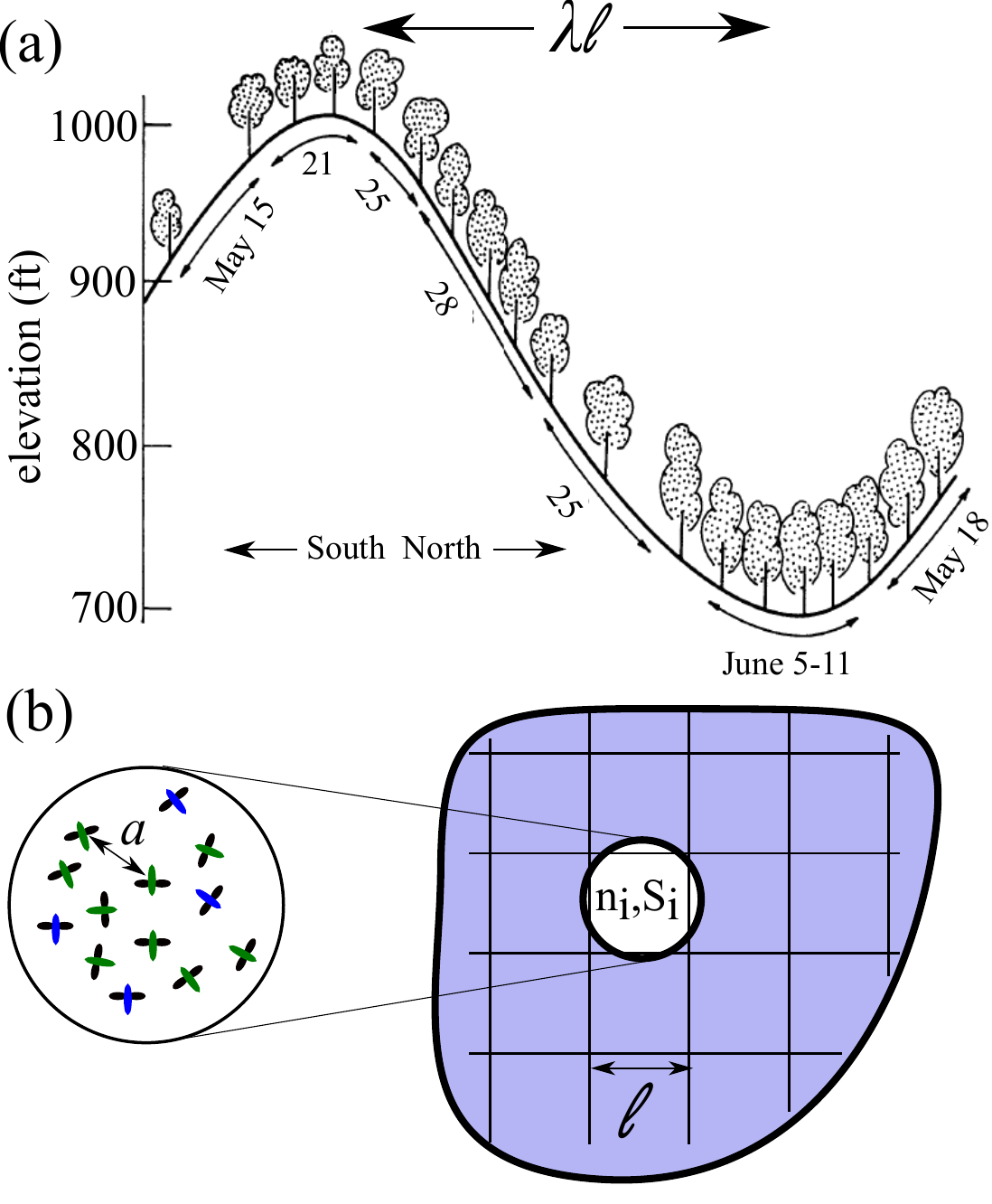}
\caption{Lateral temperature variations. (a) Topography of an Ohio forest, indicating 
forest density and dates of 
cicada emergences.  Adapted from \cite{Heath1968}.  (b) Region of burrowing nymphs, with typical spacing $a$, coarse-grained on the scale $\ell$.  Microclimates are correlated on the scale $\lambda \ell$.}
\label{fig3}
\end{figure}

{\it Microclimates and coarse-graining.}
The above does not account for {\it lateral} variations in temperature with elevation, tree cover, and solar exposure, which 
determine the local {\it microclimate}.  As Heath
showed, the days of cicada emergences varied with location in a hilly landscape as shown in Fig. \ref{fig3}(a) \cite{Heath1968}. Sunny, sparsely forested south-facing slopes
have the earliest swarms, with successive swarms typically separated by
a gap of several days, disproving the simplistic view that all cicadas in a brood emerge
at once within a few days; the entire process within an emergence year may take a month.

While a full description of microclimate requires
accounting for topography, solar exposure, and vegetation, 
we argue that the net effect of these contributions is that nymphs experience a {\it quenched, spatially
correlated random temperature field}.
From our analysis of underground temperatures, we identify
the annual penetration length $\ell_a$ as the smallest scale
of that random field which, therefore, serves as a coarse-graining length $\ell$.
The area density $n$ of cicadas can reach 
$10^6\,$/acre $\sim 250\,$/m$^2$ \cite{millions}, with
average distance $a\sim 1/\sqrt{n}$ between nymphs as small as
$5-10\,$cm $\ll \ell\sim 2$m. 
We adopt the coarse-grained representation of the population density 
$n({\bf r})$ at point ${\bf r}$ in Fig. \ref{fig3}(b), 
where each subgroup $b_i$ of 
area $\ell^2$ is associated to a site on a square lattice at location ${\bf x}_i={\bf r}_i/\ell\in \mathbb{Z}^2$ and, as in a lattice-gas description, 
is assigned an occupation variable   
$n_i$ denoting if it is empty ($0$) or occupied ($1$).

The burrowing depth of nymphs, and the separation of scales $a\ll \ell$ suggest 
that a natural model of the thermal environment of cicadas
involves a two-dimensional temperature field 
$\tau({\bf x}_i,t)=\tau_m(t)+\tilde{\tau}_f({\bf x}_i)$,
partitioned into a slowly rising mean $\tau_m(t)$ obtained from $T(z_b,t)$ in 
\eqref{subsurfacetemp} by averaging over the fast daily oscillations,
and a term $\tilde{\tau}_f({\bf x}_i)$ that encodes the fluctuations
in the local microclimate.
Shifting
the origin of temperature to be $T_c$, near the crossing day we may write
${\tau}_m(t)\simeq \tilde{\alpha}(t-t_c)$, where $\tilde{\alpha}\simeq 0.15\,^\circ$C.
We assume that $\tilde{\tau}_f({\bf x})$ is a Gaussian random field with zero mean and some
two-point correlation
\begin{equation}
    C(\vert {\bf x}_i-{\bf x}_j\vert)=\langle \tilde{\tau}_f({\bf x}_i)\tilde{\tau}_f({\bf x}_j)\rangle. 
\end{equation}
In practice we assume an exponential correlation
$C=\sigma^2 e^{-\vert{\bf x}_i-{\bf x}_j\vert/\lambda}$ with a single (scaled) length $\lambda$, where $\sigma$ is the standard
deviation of the local field, in the range $\sim 1-3\,^\circ$C.
From the topography of Fig. \ref{fig3}(a)
and contour maps of the regions studied by Heath, we deduce $\lambda \sim 50$.

\begin{figure*}[t]
\includegraphics[width=2.0\columnwidth]{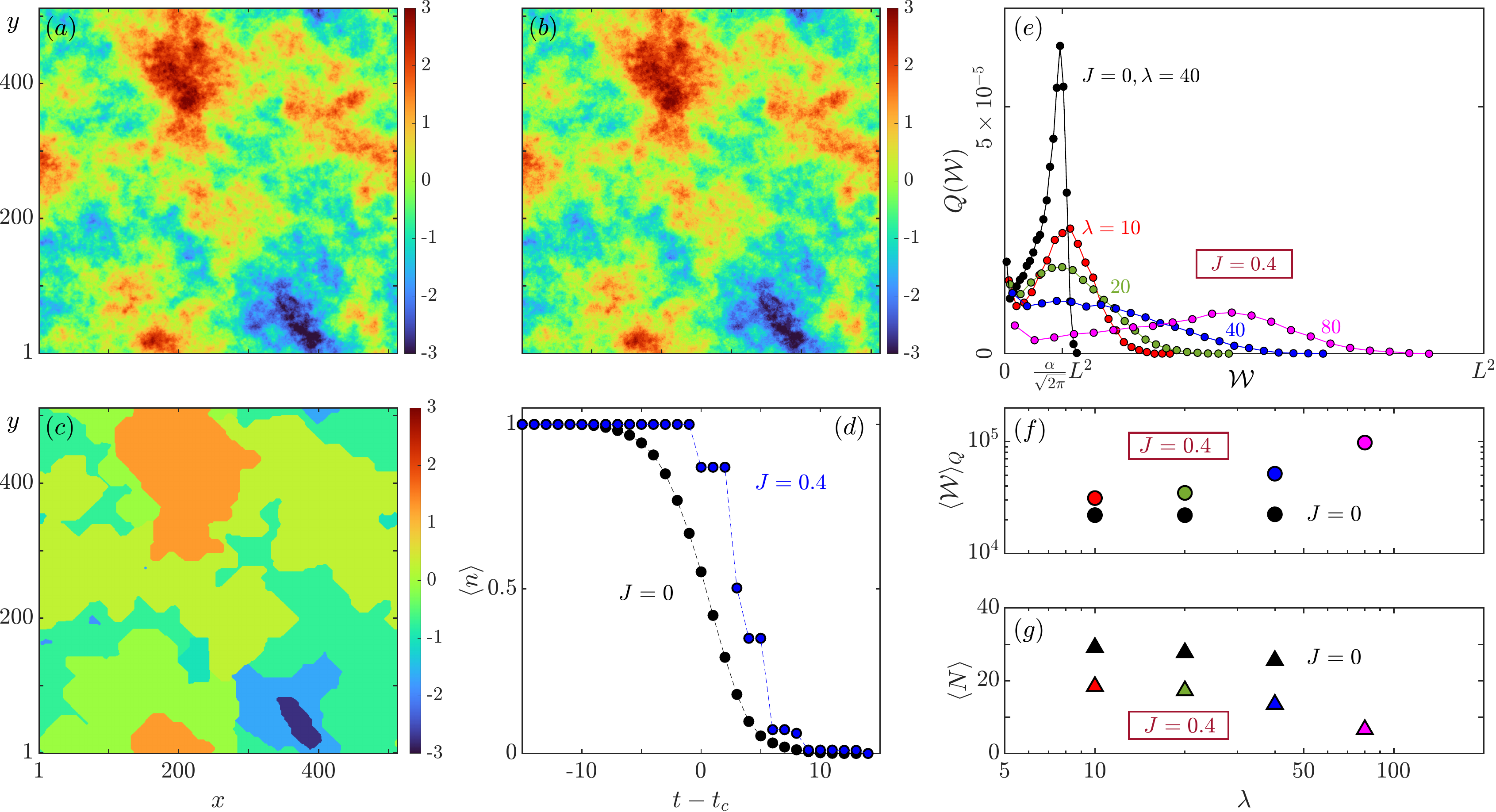}
\caption{Numerical results with $L=512$. (a) Realization of the random field $\tau_f({\bf x})$. (b,c) Composite plots of swarms for $J=0$ and $J=0.4$, respectively, 
color-coded by mean value of $\tau_f$ within each swarm, with $\alpha=0.3$.
(d) Occupancy versus time for cases in (b,c).  (e-g) Results from averaging
over $10^4$ realizations of $\tau_f$ for $J=0$ (black) and $J=0.4$ (colors).
(e) Binned swarm size distribution $Q({\cal W})$ for $J=0$ and for 
$J=0.4$ and several values of $\lambda$. (f,g) 
Average swarm size experienced by a cicada and
average number of swarms versus $\lambda$.  At $J=0$, the decrease in $\langle N\rangle$ 
for $\lambda \gtrsim 40$ is a finite-size effect.}
\label{fig4}
\end{figure*}

{\it Model of decision-making.}  To complete the model by allowing for 
nymph communication, we introduce a second variable at each 
site: a spin-like scalar $S_i(t)$ that characterizes 
the binary choice at a given time: to remain underground ($-1$) or to emerge ($+1$).
The decision of group $b_i$ to emerge is determined by the local temperature and the behaviour of other groups in the neighbourhood 
$V_i$ (the $q=8$ nearest- and next-nearest-neighbors of site $i$) via the field ${\cal H}_i(t)={\cal J}_i(t)+\tau({\bf x}_i,t)/\sigma$, where the temperature has been non-dimensionalized by $\sigma$, and
\begin{equation}
    {\cal J}_i(t)
    = J \sum_{j \in V_i}  n_j(t) S_j(t) \, ,
    \label{Ji}
\end{equation}
in which we adopt the simplest model with a single coupling $J$ throughout the neighborhood.  Hence
\begin{equation}
    {\cal H}_i (t)=\alpha\left(t-t_c\right) + \tau_f({\bf x}_i) + {\cal J}_i(t),
    \label{rescaledH}
\end{equation}
where $\alpha=\tilde\alpha/\sigma$, and $\tau_f=\tilde{\tau}_f/\sigma$ has  unit variance.
As in previous models of collective decision-making \cite{Michard2005,Bouchaud2013}, 
the decision of $S_i$ to flip from
$-1$ to $+1$ occurs when ${\cal H}_i$ becomes positive, as in the ``zero-temperature" limit of the RFIM approach. When $J=0$, each spin flips to $+1$ when its local
temperature field crosses the threshold.
When $J > 0$, a subgroup's decision to emerge is reinforced by occupied neighboring sites that have  flipped, a feature
that leads to swarms.
${\cal H}_i$ plays the same role as the local field in a spin model of magnetization;
with the occupation variables $n_i$, the system is random field Ising model (RFIM)
with annealed site dilution.
In most studies of the RFIM the random field is independent from site to site, but here the 
microclimates are correlated on scales large compared to the lattice spacing.

The dynamics of decision-making by subgroups is modelled as a discrete-time process in which state variables
are updated daily, without resolving the behavior within each day.  
In numerical studies, we start at $t<t_c$ 
with full occupancy ($n_i=1$, $\forall i$), and with all subgroups choosing to
remain underground ($S_i=-1$, $\forall i$).  On each day we iteratively 
update the spins by the rule
$S_i^{k+1}={\rm sign}\left[{\cal H}_i^k\right]$,
where $k=1,2,\ldots$ labels iterations, until no more spins flip to $+1$.  
We call a {\it swarm} the set ${\cal A}(t)$ of spins that have flipped from $-1$ to $+1$ on a given day.
The occupancy variables of sites 
in ${\cal A}(t)$ are set to zero when the updating rule is complete for that day.
The process continues on successive days until the entire lattice is empty.

{\it Numerical studies.} The model \eqref{rescaledH} has three dimensionless parameters ($\alpha$, $J$, $\lambda$) and the dimensionless system size $L$ \cite{randomfieldFourier}.
Since the tails of $\tau_f$ determine the first and last swarms, some 
$95$\% of the cicadas emerge over
a period of $4/\alpha$ days, during
which time the mean temperature sweeps from $-2\sigma$ to $+2\sigma$ of the
random field;
setting $\alpha=0.3$ spreads swarms over the realistic time of $\sim\! 14$ days.
Consider first the effect of the coupling $J$ at fixed $\lambda$. 
Figure \ref{fig4}(a) shows a realization of $\tau_f({\bf x})$ with $\lambda=30$, within which are
correlated local ``hot-spots" and ``cold-spots" in the landscape:
like sunny hilltops and shaded valleys.   If $J=0$ (Fig. 
\ref{fig4}(b)),  
the swarms are composed of those sites whose random field values fall in intervals of size $\alpha$.  
As $\tau_f$ is Gaussian, the lattice occupancy versus time (Fig. \ref{fig4}(d)) is 
a discretely-sampled error function $\langle n\rangle \approx \erfc(\alpha(t-t_c)/\sqrt{2})/2$.  In contrast, when $J>0$ (Fig. \ref{fig4}(c)), inter-cicada coupling produces large coherent domains. 
Emptying the lattice involves a smaller number of large swarms, which may be separated
by time gaps without activity, as in Heath's observations \cite{Heath1968}.  
This picture\textemdash of quiescent periods punctuated by large emergence events\textemdash 
resembles the avalanches seen in the conventional RFIM, but the event initiation
differs due to the daily resetting of the occupancy variables.

Next we examine properties of swarms averaged over $10^4$ realizations of 
$\tau_f$, through the distribution $P({\cal W})$ of swarm 
sizes ${\cal W}$, with mean $\langle {\cal W}\rangle_P=\sum  
{\cal W}P({\cal W})$ and $Q({\cal W})={\cal W}P({\cal W})/\langle {\cal W}\rangle_P$, the probability that a given cicada emerges in a swarm of size ${\cal W}$ \cite{computations}.
We see in Fig. \ref{fig4}(d) that when $J=0$ the largest swarms occur near $t_c$, where from
the form of $\langle n\rangle$ above we deduce the maximum average swarm size to be 
$\sim\!\!\alpha L^2/\sqrt{2\pi}$.
This sharp cutoff is clearly visible in 
$Q({\cal W})$ shown in Fig. \ref{fig4}(e).
In contrast, when $J=0.4$ the pdf $Q({\cal W})$
broadens with increasing correlation length 
of the random field, signifying the existence
of ever larger swarms.
This is further quantified by examining 
$\langle {\cal W}\rangle_Q=\sum 
{\cal W}Q({\cal W})$, the
average size of a swarm in which a given
cicada emerges.
Figs. \ref{fig4}(f,g) show that beyond $\lambda\sim 20$, the effect of communication ($J>0$) is that the 
average swarm size is larger, and the number of swarms depends more strongly on 
$\lambda$.  These trends continue for 
larger $J$.

We have shown that the statistical physics of collective decision-making,
quantitatively based on the thermal physics of local microclimates, reproduces key known features
of periodical cicada emergences:  compact, large swarms spread over several weeks, with
 temporal gaps between them.
Future work could focus on 
testing the hypothesis of communication between nymphs, and quantifying 
spatial variations in microclimate and their correlation with emergences.  Here we have focused on the synchrony of emergences
in year 17.  It remains to be seen whether collective decision-making can explain the 13- and 17- year synchrony.  Finally we ask: Is there a biological system that exhibits periodic emergences 
on shorter time scales, allowing for convenient study of this 
magical phenomenon?

\begin{acknowledgments}
We thank Anne Herrmann for discussions at an early stage of this work and Sumit K. 
Birwa for assistance with numerical computations.  
\end{acknowledgments}



\begin{thebibliography}{99}
\bibitem{computations} Calculations were performed
on the DAMTP HPC cluster Fawcett using a bespoke Matlab script in task-farming mode.
\end{thebibliography}
\end{document}